\documentclass{article}

\usepackage{authblk}
\usepackage[english]{babel}
\usepackage{amsthm}
\usepackage{amssymb}
\usepackage{tikz}
\usepackage{booktabs}
\usetikzlibrary{positioning, shapes.geometric, arrows.meta}

\theoremstyle{definition}

\theoremstyle{remark}

\usepackage[title]{appendix}
\usepackage{color,soul}
\usepackage{amssymb, amsmath}

\usepackage{amsmath}
\usepackage{algorithm}
\usepackage{verbatim}
\usepackage{graphicx}
\usepackage{algpseudocode}
\date{\today}

\title{Physics-Guided Inverse Regression for Crop Quality Assessment}

\author[1]{David Shulman}
\author[2,3]{Assaf Israeli}
\author[3]{Yael Botnaro}
\author[3]{Ori Margalit}
\author[3]{Oved Tamir}
\author[3]{Shaul Naschitz}
\author[3,4]{Dan Gamrasni}
\author[3,4]{Ofer M.~Shir}
\author[1]{Itai Dattner}
\affil[1]{Department of Statistics, University of Haifa, Haifa, Israel}
\affil[2]{The Hebrew University of Jerusalem, Israel}
\affil[3]{Migal Institute, Qiryat Shmona, Israel}
\affil[4]{Tel-Hai College, Upper Galilee, Israel}

\begin{document}

\maketitle

\begin{abstract}
We present an innovative approach leveraging Physics-Guided Neural Networks (PGNNs) for enhancing agricultural quality assessments. Central to our methodology is the application of physics-guided inverse regression, a technique that significantly improves the model's ability to precisely predict quality metrics of crops. This approach directly addresses the challenges of scalability, speed, and practicality that traditional assessment methods face. By integrating physical principles, notably Fick’s second law of diffusion, into neural network architectures, our developed PGNN model achieves a notable advancement in enhancing both the interpretability and accuracy of assessments. Empirical validation conducted on cucumbers and mushrooms demonstrates the superior capability of our model in outperforming conventional computer vision techniques in postharvest quality evaluation. This underscores our contribution as a scalable and efficient solution to the pressing demands of global food supply challenges.

\end{abstract}

\begin{keywords}
Physics-Guided Neural Networks (PGNNs); Machine Learning in Agriculture; Agricultural Quality Assessment; Moisture Distribution Modeling; Inverse Regression
\end{keywords}


\section{Introduction}
\subsection{Background and motivation}
The imperative for advancing quality assessment methods in agriculture is underscored by the critical role these assessments play in supply chain management, impacting economic and consumer health outcomes \cite{kader2004role,kiaya2014post,rajapaksha2021reducing,wognum2011systems}. Traditional approaches, predominantly based on laboratory measurements, face limitations in terms of scalability, speed, and practicality, especially in varied agricultural contexts \cite{gomiero2018food}. These methods are often labor-intensive and require extensive equipment, rendering them less effective for broad application.

The motivation for our work is rooted in enhancing postharvest technologies, which are pivotal in global agro/food industries. These technologies become increasingly crucial in light of the growing demand for affordable food supplies amidst a burgeoning world population \cite{fao2018future}. Postharvest losses remain a significant challenge, accounting for up to 20\% losses of the total produce in well-developed food systems and potentially more than 30\% in less developed systems \cite{spang2019food}. 

Emerging technologies of computer vision for image processing and machine learning, offer efficient alternatives for quality assessment, allowing rapid and non-destructive evaluation of produce quality through advanced image analysis \cite{bhargava2021fruits,borras2015data,he2016deep,lecun2015deep,goodfellow2016deep}. Deep learning \cite{lecun2015deep,goodfellow2016deep}, particularly deep convolutional neural networks (CNNs) \cite{krizhevsky2012imagenet}, has revolutionized the way we approach image classification tasks. Their success underscored the potential of deep learning in handling complex image classification tasks, setting the stage for subsequent advancements in agriculture \cite{cho2021determination}. Recently, there has been increasing attention on the theoretical foundations of these methods \cite{schmidt2020nonparametric}. The authors of \cite{kohler2022rate} delve into the intricacies of CNNs, investigating the rate of convergence of image classifiers, while \cite{langer2022statistical} present a statistical analysis tailored to image classification problems, offering insights into the underlying mechanics of CNNs and their performance metrics. The work of  \cite{kohler2021over} critically examine over-parameterized deep neural networks, particularly their tendency to minimize empirical risk without guaranteeing generalization on new, unseen data. This observation serves as a cautionary note for practitioners, emphasizing the need for a nuanced understanding of model complexity and its impact on generalizability. Moreover, these methods often require substantial annotated data and have been criticized for their lack of transparency, being termed 'black box' systems.

In recent years, the integration of physical principles into neural network architectures has emerged as a promising approach in various scientific and engineering domains. This introduction outlines two such approaches: Physics-Informed Neural Networks (PINNs) \cite{raissi2019physics, karniadakis2021physics, cuomo2022scientific} and Physics-Guided Neural Networks (PGNNs) \cite{karpatne2017theory,daw2022physics,9126036}.

PINNs represent a novel methodology where neural networks are designed to solve differential equations by incorporating physical laws directly into the network's loss function. This integration ensures that the network's predictions adhere to established physical principles. Commonly utilized in solving both forward and inverse problems in physics and engineering, PINNs offer solutions where traditional methods might struggle, particularly in cases where the governing equations are known, but solutions are complex.

In PGNNs, physical knowledge is embedded within the network, especially in the input and hidden layers, often through a hybrid architecture that melds traditional physics-based models with neural network components. This approach emphasizes the embedding of prior physical knowledge into the network, which not only guides the learning process but also enhances the network’s capability to learn and replicate the underlying physical laws effectively. PGNNs are particularly useful in scenarios where some physical knowledge is available and can be leveraged to improve the network's accuracy and reduce reliance on extensive training data. 

Both PINNs and PGNNs represent significant strides in the field of computational physics and machine learning, offering advanced frameworks for integrating domain-specific knowledge into data-driven models.

Building on the principles of PGNNs, we introduce a neural network model tailored for the analysis of agricultural data. This model is distinguished by its specialized approach to integrating physical laws into the neural network architecture, offering a novel perspective in agricultural data analysis.




\subsection{Overview of our approach}
The quality attribute of agricultural produce, represented by $x$, is intrinsically
linked to the resultant quality metric $y$. Specifically, our methodology, at its core, employs a two-step physics-guided approach to estimate laboratory measurements \(Y\) from image data \(Z\). The first step involves a \textit{Physics-Guided Inverse Regression} (\(Z \rightarrow \hat{x}\)), where we harness physical laws or principles to transform image data into a representation of basic feature $x$ that encapsulate the intrinsic quality attributes of the subject, such as agricultural produce. This transformation is achieved without direct supervision, relying solely on the inherent information in \(Z\) and the governing physical laws. This process not only aims to capture the physical essence embedded within the image data but also to distill it into quantifiable parameters that reflect the underlying quality attributes.

Following the extraction of these basic physical parameters, our approach advances to \textit{Predictive Modeling} (\(\hat{x}\rightarrow Y\)), utilizing the estimated features \(\hat{x}\) to predict laboratory measurements \(Y\). This phase of the methodology can employ a variety of modeling techniques, ranging from regression models to advanced machine learning and deep learning frameworks, depending on the complexity and nature of the relationship between $x$ and $y$. 


Inverse imaging problems, particularly in fields such as medical imaging \cite{ongie2020deep,yaman2020self}, often employ PGNN to reconstruct images $x$ from observed data $y$ based on the equation \(y=Ax\), utilizing various physical properties. While these approaches have demonstrated significant success, they typically focus on a one-directional reconstruction of $x$ from $y$.

Our method distinguishes itself by employing a two-step predictive modeling strategy, where the initial physics-guided transformation from image data \(Z\) to basic features \(\hat{x}\) is followed by predictive modeling to estimate laboratory measurements \(Y\). This approach not only addresses the challenge of directly predicting $Y$ from $Z$ but also enriches the interpretive depth of the model by grounding the prediction in physically derived features $x$. This innovative strategy opens new avenues for leveraging PGNN beyond direct target prediction, facilitating a deeper understanding and exploration of the underlying physical mechanisms that influence observable outcomes.

The advent of sophisticated statistical methods and computational techniques has significantly advanced the field of image processing and analysis \cite{fieguth2010statistical}. The aforementioned book specifically  a specific attention is given to the inverse problem that we address in this work in an unsupervise manner. Indeed, in the quest for models that better mimic human learning, the role of unsupervised learning in deep convolutional neural networks has gained attention \cite{rawat2017deep}.

Focusing on cucumbers and mushrooms as specific case studies, we apply PGNN to postharvest analysis. Cucumbers and mushrooms are known for their sensitivity and substantial postharvest losses \cite{cortbaoui2015new}. They undergo rapid quality degradation during storage and marketing due to factors such as ethylene sensitivity, high water content, and susceptibility to chilling injuries and rots \cite{poenicke1977ethylene, manjunatha2014effect, nasef2018short}. The freshness and overall quality of cucumbers are often influenced by their moisture content. As cucumbers or mushrooms age or are stored under suboptimal conditions, they tend to lose moisture, leading to a decline in quality. Accurate moisture modeling, for which we employ Fick's second law in our algorithm, is therefore crucial \cite{cantor2020equations,kahramanouglu2019improving,omolola2017quality,yahia2018postharvest,onwude2016modeling}.

The effectiveness of PGNN in cucumber postharvest quality assessment is validated through a comprehensive case study. By integrating Monte Carlo simulations with real-world data analysis, we demonstrate the model's adaptability and superior performance in comparison to established computer vision standards, like the ResNet model \cite{he2016deep}.


The paper is structured as follows: Section 2 details the development of our statistical model and neural network architecture. Section 3 outlines the Monte Carlo simulation setup. Section 4 presents the application of our methodologies to  cucumber and mushrooms freshness analysis. Section 5 concludes with a summary of our findings and a discussion on their broader implications, including potential areas for future research.

\section{Methodology}
\subsection{Problem Formulation}

To simplify the presentation and maintain generality in our formulation, we omit the explicit specification of each variable's dimensions. However, in subsequent sections where the theoretical framework is applied to a specific context, we will provide the dimensions of all involved quantities.
The quality attribute of agricultural produce, represented by \( x \), is intrinsically linked to the resultant quality metric \( y \). The relationship between these variables is formalized as:
\[
y = f(x),
\]
where \( f \) is a function mapping square-integrable functions to real numbers. In the context of cucumbers, for example, laboratory measurements, denoted by \( Y \), essentially reflect the underlying physical states of the produce through an unknown function \( f \).

For the laboratory measurements $Y$, we propose the following measurement error model:
\begin{equation}\label{eq:MEmodel}
Y_i = f(x_i) + \eta_i, \quad i = 1,...,n,
\end{equation}
where $Y_i$ denotes the laboratory measurement for the ith sample. In this model, $f$ represents the true relationship between the quality attribute $x$ and the laboratory measurement, while 
$\eta_i$ is an error term accounting for noise in the measurements. 

Furthermore, we propose that the image data \( z \), capturing the visual attributes of the produce, can be modeled as:
\begin{equation}\label{eq:g}
z = g(x),
\end{equation}
where \( g \) represents a transformation mapping \( x \) into its visual representation.

In our model, \( x \) serves as the fundamental driving force behind the observed data, encapsulating the inherent dynamical features that are responsible for the characteristics observed in both laboratory measurements and image capturing.

The process of image acquisition is described by:
\begin{equation}\label{eq:obsz}
Z_i = g(x_i) + \epsilon_i, \quad i = 1,...,n,
\end{equation}
where \( \epsilon_i \) are noise variables.

The primary objective is to utilize our dynamic model for \( x \) along with the data \( (Z_i,Y_i), i=1,...,n \), to effectively learn \( f \). This approach integrates theoretical models of \( x \)'s behavior with empirical observations, enhancing our understanding and prediction capabilities. 


\subsection{Integration of Physical Models in Quality Attribute Analysis}

Our study examines a particular quality attribute of agricultural produce, denoted by \( x \). This attribute adheres to specific physical laws that govern its evolution both spatially and temporally. Such adherence ensures that \( x \) is a reliable indicator of the produce's quality, influenced by its physical environment.

\paragraph{Modeling Framework:} To effectively capture the dynamics of \( x \), we utilize a differential equation framework that aligns with the physical laws influencing its evolution. This is represented as follows:
\begin{equation}\label{diff_operator}
\frac{\partial x}{\partial t} = \mathcal{P}(x, \nabla x, \nabla^2 x, \ldots),
\end{equation}
where \( \mathcal{P} \) is a differential operator representing the combined effect of the physical laws governing \( x \). This operator may include, but is not limited to, aspects of diffusion, advection, reaction, and other relevant physical phenomena, depending on the specific nature of \( x \) and the agricultural produce under study.

The choice of \( \mathcal{P} \) is guided by the physical characteristics of the produce and the specific quality attributes being modeled. The equation above serves as a general framework, which can be specified further based on empirical data and theoretical insights into the physical processes affecting \( x \).

\subsection{Conceptual Framework for Physically-Informed Mapping}

At the heart of our approach is the relationship between observable image data, denoted as \( z \), which captures the visual attributes of agricultural produce, and the intrinsic quality attribute, represented as \( x \). We introduce a conceptual framework that establishes a mapping informed by physical laws, connecting these two elements.

\textbf{Methodological Approach - Mapping from \( z \) to \( x \)}: Our methodology centers around the function \( {g}^{-1} \), the inverse of \( g \) given in Eq. (\ref{eq:g}). This function \( {g}^{-1} \) maps the image \( z \) back to the quality attribute \( x \).  The mapping through \( {g}^{-1} \) is deeply informed by the physical laws that influence the produce's characteristics. It adeptly converts the visual and textural features present in \( z \) into quantifiable and meaningful indicators of quality \( x \), effectively reversing the mapping process of \( g \). This ensures a precise translation of the characteristics observed in the image data into their corresponding physical attributes.

This framework underpins the development of models and algorithms that leverage image data for estimating and predicting the quality of agricultural produce. It underscores the crucial insight that the visual characteristics in \( z \) are not superficial but are intimately linked to the deeper, physically-grounded qualities denoted by \( x \). The mapping function \( {g}^{-1} \) thus plays a vital role in translating visual information into tangible physical qualities, enabling a nuanced and accurate assessment of agricultural produce quality.
\subsection*{Formulation of the Optimization Problem}
Prior to delving into the specifics of our optimization challenge, it is important to outline the preliminary phase where our physics-guided framework is trained to map image data \( Z_i \) given by Eq. (\ref{eq:obsz}) to estimated quality attributes \( \hat{x}_i \). This process inherently involves hypothesizing an optimization over the space of possible inverse functions \( \hat{g}^{-1} \), guided by the physical laws pertinent to the system under study. Unlike traditional methods that heavily rely on labeled data for training, our model uniquely capitalizes on the input data \( Z_i \) alone, drawing on physical laws to inform the learning mechanism. The goal is to refine \( \hat{g}^{-1} \) such that the estimated quality attribute \( \hat{x} \) closely approximates the true quality attribute \( x \), within the bounds of an acceptable discrepancy \( d(\hat{x}, x) \) defined by a threshold \( \delta \). This expectation is predicated on the availability of a sufficiently diverse dataset that encapsulates the variability intrinsic to the physical phenomena being modeled.

The optimization challenge, now formally introduced, is articulated as follows:
\begin{equation}\label{eq:optimization_challenge}
    \min_{\hat{g}^{-1}} \sum_{i=1}^{n} \mathcal{L}_{physical}( \hat{g}^{-1} (Z_i)),
\end{equation}
where \( \mathcal{L}_{physical} \) denotes a physics-guided loss function designed to evaluate the extent to which the predictions of \( \hat{g}^{-1}(Z)\) adhere to the established physical model. This assessment hinges on criteria directly extracted from the governing physical laws. The act of minimizing over \( \hat{g}^{-1} \) entails adjusting the parameters of \( \hat{g}^{-1} \) to reduce the loss function \( \mathcal{L}_{physical} \), thereby ensuring that the estimated outputs align more closely with the physical model. Constraints in this optimization process may include ensuring the feasibility of solutions within the physical and computational limits, maintaining the integrity of the physical laws in the predictions, and possibly incorporating regularization terms to prevent overfitting.

By structuring our models around this optimization framework, we create physics-constrained models that are not only innovative in their independence from labeled data but also in their fidelity to the physical insights that drive their predictive capabilities.

\subsection{Predictive Modeling Based on Image Data for Laboratory Measurements}

Our methodology involves predicting quality measure \( Y \) as given in Eq. (\ref{eq:MEmodel}) from image data \( Z \) given by (\ref{eq:obsz}). We employ two principal strategies for this prediction:

\paragraph{Method 1: Direct Prediction}
The first strategy is to develop a predictive model, designated as \( \hat{h} \), that establishes a direct link between image data \( Z \) and quality measurements \( Y \).

\textbf{Model \( \hat{h} \) - Direct Prediction:} Model \( \hat{h} \) is configured to take image data \( Z \) as input and output an estimated value for the quality measurement \( Y \). This model is constructed using a dataset containing corresponding instances of \( Z \) and \( Y \).

This direct prediction method is efficient and effective when a strong, direct correlation exists between the image data and the laboratory measurements. However, it may not fully exploit the underlying physical principles affecting the produce's quality.

\paragraph{Method 2: Inverse Prediction}
The second approach utilizes the Physics-Guided Mapping Function \( \hat{g}^{-1} \) to first estimate the quality attribute \( \hat{x} \) via Eq. (\ref{eq:optimization_challenge}) from the image data \( Z_i \). Subsequently, a separate model, \( \hat{f} \), is applied to predict quality measurements \( Y_i \) based on the estimated quality attribute \( \hat{x} \).

\textbf{Model \( \hat{f} \) - Inverse Prediction:} Model \( \hat{f} \) uses the estimated quality attribute \( \hat{x} \) as input to predict the quality measurement \( Y \). Training for this model is carried out on a dataset where estimated values \( \hat{x} \) (derived from \( Z \)) and actual measurements \( Y \) are paired.

\paragraph{Effectiveness of the Inverse Prediction}
The inverse prediction approach relies on two critical aspects:
\begin{itemize}
    \item The accuracy of \( \hat{g}^{-1}: Z \rightarrow \hat{x} \) in estimating the true quality attribute \( x \), measured by the metric \( d(\hat{x}, x) \).
    \item The precision of \( \hat{f}: \hat{x} \rightarrow Y \) in mapping the estimated quality attribute \( \hat{x} \) to the quality measurements \( Y \), evaluated using the metric \( d(\hat{Y}, Y) \).
\end{itemize}

We hypothesize that this two-stage inverse prediction method might yield more accurate predictions than the direct approach \( h: Z \rightarrow Y \). This is particularly likely when the combined inaccuracies of the two-stage process (\( d(\hat{x}, x) \) and \( d(\hat{Y}, Y) \)) are smaller than the error in the direct model \( h \). The inverse method is especially advantageous in scenarios where the quality attribute \( \hat{x} \) includes complex characteristics of the produce not directly observable in the image data \( Z \), thus enabling a more detailed and physically informed prediction.

\subsection{Methodological Adaptation for Vegetable Images}
\subsubsection{Overview}
In this section, we demonstrate the practical application of our theoretical framework by analyzing vegetable images for quality assessment. This case study involves linking laboratory measurements with image data and understanding the moisture content changes within vegetables, guided by Fick's second law of diffusion.

We chose neural networks (NNs) for our mapping functions due to their proficiency in handling non-linear data patterns. Their ability to learn complex, non-linear relationships is crucial in accurately modeling the intricate behaviors in agricultural image data. This makes NNs particularly suitable for our application, where the relationships between image features and quality attributes are often non-linear and dynamic.

\subsubsection{Definitions and Data Description}

\paragraph{Laboratory Measurements of Quality (\( Y \))}
For each vegetable sample, laboratory measurements, denoted as a scalar \( Y \), include various metrics such as elastic resistance, weight change, luminosity, and size change.

\paragraph{Image Acquisition (\( Z \))}
The image dataset comprises RGB images of the vegetables. These images capture critical visual attributes such as color, texture, and shape and of size $N\times M$. 

\paragraph{Quality Attribute (\( x \))}
The moisture content of the vegetables is considered as the quality attribute of the agricultural produce, represented by \( x \).
\paragraph{Moisture Content and Physical Laws}
The spatial and temporal variations in moisture content, denoted as \( x \), are modeled within the confines of a diffusion equation. This modeling approach is a specific instantiation of the broader differential equation framework outlined in Eq. (\ref{diff_operator}). In this context, we apply Fick's second law of diffusion, tailored to a two-dimensional Cartesian coordinate system defined by the coordinates \( u \) and \( v \). This adaptation allows for a precise representation of moisture diffusion within agricultural products, capturing the essential dynamics governed by the physical law.




Fick's second law, expressed in two-dimensional Cartesian coordinates \((u, v)\), is given by:
\begin{equation}\label{diff_operator_cartes}
\frac{\partial x}{\partial t} = D \left( \frac{\partial^2 x}{\partial u^2} + \frac{\partial^2 x}{\partial v^2} \right),
\end{equation}
where \( x(u, v, t) \) denotes the moisture concentration as a function of spatial coordinates \( u \), \( v \), and time \( t \). The diffusion coefficient \( D \) characterizes the rate of moisture dispersion through the vegetable tissue.

This physical law, particularly its spatial aspects, is instrumental in estimating the inverse function  \( {g}^{-1} \). By integrating Fick's law into the modeling process, \( \hat{g}^{-1} \) is designed to accurately reflect the moisture dynamics within vegetables. This approach ensures that the mapping from image data to moisture content not only adheres to empirical data but is also grounded in the physical principles governing moisture diffusion, thus enhancing the reliability and accuracy of the predictions made by \( \hat{g}^{-1} \).

\begin{figure}[!ht]
\centering
\includegraphics[width=1\textwidth]{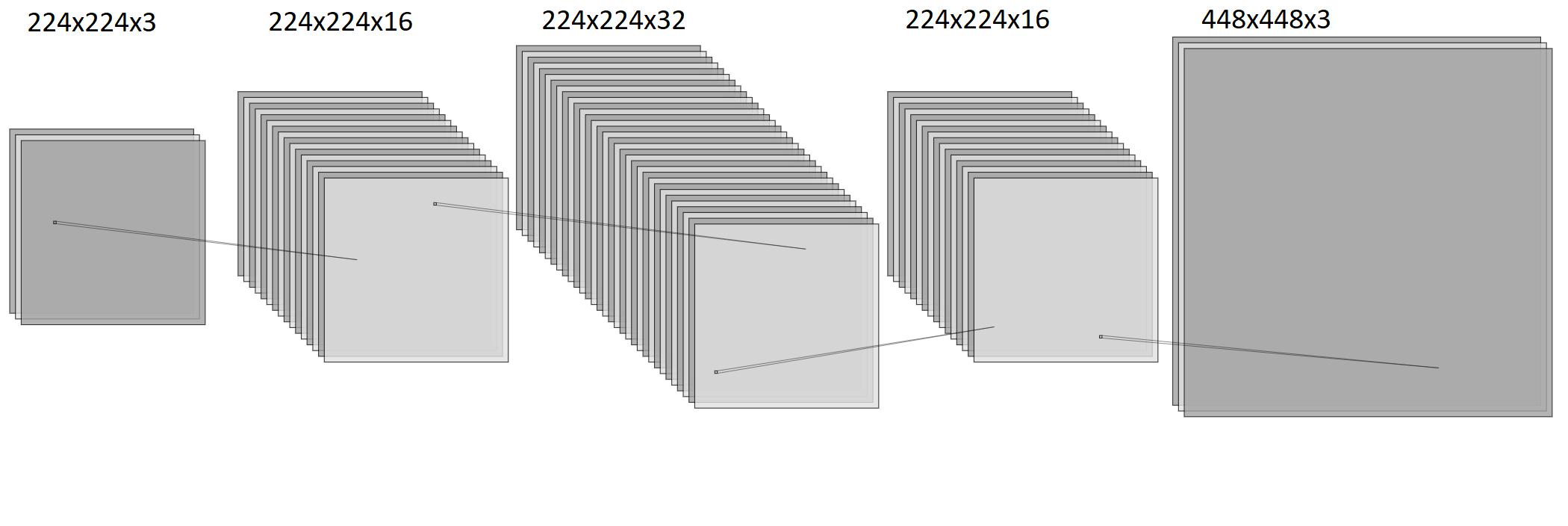}
\caption{The Physics-Guided Mapping Function \( \hat{g}^{-1} \) Architecture.}
    \label{fig:cnn}
\end{figure}
\subsubsection{Neural Network Architectures for Mapping Functions}

In our approach, we utilize three distinct neural network models for the mapping functions, each tailored to specific prediction strategies.

\paragraph{Model \( \hat{h} \) and \( \hat{f} \) - ResNet-18 Based Architectures}
Both \( \hat{h} \) for direct prediction and \( \hat{f} \) for the second stage of inverse prediction share a similar architecture based on ResNet-18, a deep convolutional neural network. ResNet-18, part of the Residual Network family, is known for its ability to effectively handle deep learning tasks without suffering from performance degradation due to increased network depth. The core idea of ResNet-18 is to introduce skip connections (or shortcuts) to allow the flow of gradients directly through the network, mitigating the vanishing gradient problem.  ResNet-18, with its 18 layers, provides a good balance between complexity and performance for our application.

To tailor the ResNet-18 architecture for our specific regression problem, we acknowledge that ResNet-18 outputs 512 features by design. This output is not directly suited for regression tasks without additional adjustments. Therefore, we enhance the architecture with a fusion block comprising two fully connected  layers. This modification serves a dual purpose: for \( \hat{h} \), it enables the direct mapping of image data \( Z \) to the corresponding laboratory measurements \( Y \); for \( \hat{f} \), it maps an intermediate quality attribute \( \hat{x} \), estimated from the image data, to the laboratory measurements \( Y \). Through the integration of this fusion block, we leverage the powerful feature extraction capabilities of ResNet-18 and adapt it for precise, task-specific outcomes in our regression problem.


\paragraph{Model \( \hat{g}^{-1} \) - CNN Transformation Network}
The Physics-Guided Mapping Function \( \hat{g}^{-1} \), used in the first stage of inverse prediction, is an CNN Transformation Network. This model is specifically designed for the inverse regression task, mapping 3-channel image data \( Z \) to a 3-channel output representing the estimated quality attribute \( \hat{x} \). Its architecture can be summarized as follows:

\begin{itemize}
    \item Convolutional layers: Sequential layers with LeakyReLU activation and Batch Normalization. These layers, denoted as \( \text{Conv2d}_{16} \), \( \text{Conv2d}_{32} \), and \( \text{Conv2d}_{16} \), progressively capture spatial features of the image. The numbers indicate the output channels for each layer.
    \item Dropout: Applied after each convolutional layer with a dropout rate of 0.2, for regularization.
    \item Transposed convolution: A layer \( \text{ConvTranspose2d}_3 \) that upscales the feature maps to reconstruct the output image, where '3' indicates the output channels.
\end{itemize}


The CNN Transformation Network, denoted as \( \hat{g}^{-1} \), is engineered to directly transform input image data into an estimated  quality attribute \( \hat x \), ensuring adherence to underlying physical laws. This network leverages convolutional layers to extract and interpret features without spatial reduction, enabling a faithful representation of the input data. The primary objective of this architecture is to facilitate an accurate estimation of quality attributes from image data, reflecting the network's capacity to capture and apply physical principles within its predictive framework. The architecture and functionality of \( \hat{g}^{-1} \) in achieving this transformation are detailed in Figure \ref{fig:cnn}.

\subsubsection{Incorporating Steady-State Diffusion Constraints in Neural Network Modeling}

In the context of our vegetable image analysis, we consider the dataset as comprising snapshots of moisture content at a single time point. Under this scenario, we make an important assumption: the steady-state diffusion condition is applicable across the spatial domain of the images. This assumption implies that the spatial rate of change in moisture content is zero, aligning with a simplified form of Fick's second law of diffusion in steady-state.

\subsubsection*{Loss Function Incorporating Steady-State Diffusion}
Under the assumption of steady-state diffusion, the neural network \( NN_x \), which serves as the Physics-Guided Inverse Regression \( \hat{g}^{-1} \), is tasked with estimating the moisture content \( x \) from the image data \( Z \). It is essential that \( NN_x \) adheres to this physical constraint in its learning process. The loss function, designed to uphold the steady-state diffusion condition, is formulated as follows:
\begin{equation*}
\mathcal{L}_{x,n}(\theta_x) = \frac{1}{n}\sum_{i=1}^n \left( \frac{\partial^2 NN_x(Z_i; \theta_x)}{\partial u^2} + \frac{\partial^2 NN_x(Z_i; \theta_x)}{\partial v^2} \right)^2.
\end{equation*}

In this formulation, the term inside the absolute value calculates the sum of the second partial derivatives of \( NN_x \)'s output with respect to the spatial coordinates \( u \) and \( v \), representing the steady-state diffusion equation. Minimizing this loss function ensures that \( NN_x \), or \( \hat{g}^{-1} \), accurately estimates moisture content in line with the steady-state diffusion paradigm.

This modeling strategy is vital for aligning the neural network's objectives with the physical characteristics observed in the dataset. Our approach, based on the assumption of steady-state diffusion, is a suitable simplification for our dataset, which captures the spatial moisture distribution at a singular time point. By focusing on this loss function, we ensure that the outputs of \( NN_x \) (namely, \( \hat{g}^{-1} \)) are not only driven by data but also conform to the fundamental physical laws governing moisture movement within the vegetables.

\subsubsection*{Loss Function for Quality Metric Prediction}
In the second method, termed Inverse Prediction, we utilize the Physics-Guided Mapping Function \( \hat{g}^{-1} \) to initially estimate the quality attribute \( \hat{x} \) from the image data \( Z \). Following this estimation, a separate model \( \hat{f} \), represented by the neural network \( NN_y \), is employed to predict the laboratory measurements \( Y \) based on \( \hat{x} \). The parameters of \( NN_y \) are denoted as \(\theta_y\). The loss function for \( NN_y \), reflecting its accuracy in predicting these quality metrics, is structured as follows:
\begin{equation*}
\mathcal{L}_{y,n}(\theta_y) = \frac{1}{n}\sum_{i=1}^n\left( Y_i - NN_y(NN_x(Z_i;\hat\theta_x); \theta_y) \right)^2.
\end{equation*}
The loss \( \mathcal{L}_{y,n} \) measures the deviation between the actual laboratory measurements \( Y \) and the predictions by \( NN_y \), using the estimated quality attribute \( \hat{x} \), derived from \( \hat{g}^{-1} \), as an intermediary. Optimizing this loss function is crucial for ensuring the accuracy and relevance of \( NN_y \)'s predictions in the practical quality assessment of vegetables, effectively coupling the image-derived estimations with laboratory-validated measurements.

\section{Simulation}

\subsection{Motivation}
This simulation study is primarily motivated by the need to evaluate the performance of the PGNN algorithm in the context of agricultural analysis. Our focus is on modeling moisture diffusion in vegetables, a critical aspect of crop quality assessment. The study aims to assess the robustness of the PGNN model under a variety of simulated conditions, mirroring real-world agricultural scenarios. By exploring the model's behavior across different data distributions and training set sizes, we aim to gain insights into its adaptability and potential utility in practical agricultural settings. This comprehensive evaluation will contribute to a deeper understanding of the PGNN model's capabilities in addressing the complex challenges faced in agricultural data analysis.

\subsection{Data Generating Process}
\paragraph{Monte Carlo Simulation and Dataset Generation}
Utilizing a Monte Carlo approach \cite{metropolis1949monte}, each simulation iteration generates a new test set containing up to 300 samples, allowing for an extensive evaluation of the model's performance across different data distributions. This process is crucial in highlighting the model's ability to handle variability and provide reliable predictions. Training sets of varying sizes, ranging from 15 to 100 samples, are used to investigate the influence of training data volume on the model's effectiveness. This varied approach ensures a thorough examination of how the model adapts to different amounts of data, reflecting its practical utility in real-world applications.

\paragraph{Simulation Setup and Synthetic Image Generation}
The core of the simulation is based on the diffusion equation Eq. (\ref{diff_operator_cartes}), which models the spatial and temporal changes in moisture content within the vegetables. We adopt an approximate analytical solution using Fourier series \cite{bracewell1966fourier} to accommodate the boundary conditions and initial distribution of moisture. This method is particularly suited to capturing the complex dynamics of moisture movement within the vegetables, providing a realistic representation of the diffusion process. 

To generate synthetic images that reflect these dynamics, we follow a structured process:
 \begin{enumerate}
 \item Random boundary conditions and diffusion coefficients are selected to create a moisture distribution across the domain. 
 \item The total moisture content for the current domain is calculated and designated as the label for each generated instance. Subsequently, random noise is introduced to the domain's moisture content to simulate realistic variations and imperfections.
 \item A custom colormap is then applied to translate the moisture content into RGB images, enabling the direct visual representation of the moisture distribution, ensuring that the synthetic images closely mimic real agricultural scenarios while adhering to the underlying physical laws.
 \end{enumerate}

\paragraph{Quality Measure and Label Generation}
The assessment of cucumber quality in our simulation is primarily based on the total moisture content, a critical factor influencing freshness and overall appeal. This is quantified mathematically by the integral of moisture content \( x(s) \) over the spatial dimensions of the cucumber:
\begin{equation*}
    y = \int x(s) \, ds.
\end{equation*}
To account for measurement variability akin to that observed in laboratory settings, Gaussian measurement errors are introduced, forming the basis of our simulated labels \( Y\).

It is important to note, however, that while total moisture content serves as a key quality indicator in our model, this metric may not fully capture the comprehensive quality profile of cucumbers as perceived in real-world conditions. Various other factors, including but not limited to surface appearance, firmness, and absence of defects, also play crucial roles in determining the overall quality of cucumbers. Thus, our simulation's focus on moisture content should be viewed as a focused exploration of one significant aspect of quality, rather than an exhaustive measure of quality in its entirety.


\paragraph{Transformation Function \( g \) and Custom Colormap}
The implementation of function \( g \) utilizes a custom colormap strategy to translate varying levels of moisture content into colors. Specifically, we define a linear gradient between two colors—yellow for lower moisture levels and dark green for higher moisture levels. This choice is motivated by the natural visual cues associated with moisture in agricultural products.

The mathematical foundation of this mapping process is described as follows:
\[
\begin{aligned}
R(x) &= \alpha_R x + \beta_R, \\
G(x) &= \alpha_G x + \beta_G, \\
B(x) &= \alpha_B x + \beta_B,
\end{aligned}
\]
where \( R(x) \), \( G(x) \), and \( B(x) \) denote the interpolated red, green, and blue color components for a given moisture content \( x \), respectively. The coefficients \( \alpha_R, \alpha_G, \alpha_B \) along with \( \beta_R, \beta_G, \beta_B \) are meticulously chosen to ensure a seamless color gradient from yellow to dark green as the moisture level ascends.

This interpolation framework enables \( g \) to produce an RGB image where each pixel's color directly corresponds to the local moisture content, thereby offering a visual map of moisture distribution across the agricultural product. By employing this custom colormap, function \( g \) ensures that the RGB representation \( z \) of moisture content is both scientifically accurate and visually interpretable. 

\begin{figure}[ht]
\centering
\includegraphics[width=0.8\textwidth]{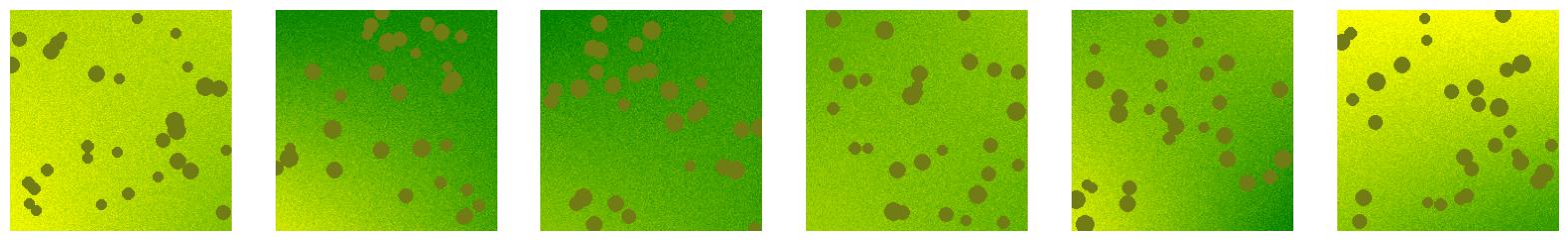}
\caption{Synthetic images generated using the moisture distribution model with the introduction of random circles as noise. Each image showcases the variability and randomness of the circle noise, simulating potential real-world imperfections or anomalies in the data.}
    \label{Simulation_images}
\end{figure}

\paragraph{Simulation Parameters and Image Processing}
Random parameters are introduced to simulate real-world conditions, including random diffusion coefficients and boundary conditions representing moisture loss. Gaussian noise is added to the simulated data to mimic real measurement imperfections, enhancing the realism of the simulation. A specific function is used to add circular imperfections to the images, see Fig. \ref{Simulation_images}. This function randomly places a predefined number of circles of varying radii on the image, effectively simulating the types of visual noise that might be encountered in real-world imaging scenarios. The aim is to assess the model's capability to handle these imperfections, which are characteristic of actual agricultural imaging environments. Image augmentation techniques and a custom color mapping strategy are employed to transform the simulated moisture content data into RGB images. These steps are instrumental in creating a dataset that closely resembles actual vegetable images, both in appearance and underlying physical properties.

\subsection{Results of the Simulation Study}

Our simulation study presents a comparative analysis of the performance of two models, ResNet and PGNN, in estimating the quality of cucumbers. The evaluation is based on several metrics: Average Root Mean Square Error (RMSE), Average Mean Absolute Error (MAE), Average Standard Errors, and Average R\(^2\) score. The results are summarized for different training set sizes, as shown in the Fig. \ref{Simulation1_a}:

\begin{figure}[!ht]
\centering
\includegraphics[width=1\textwidth]{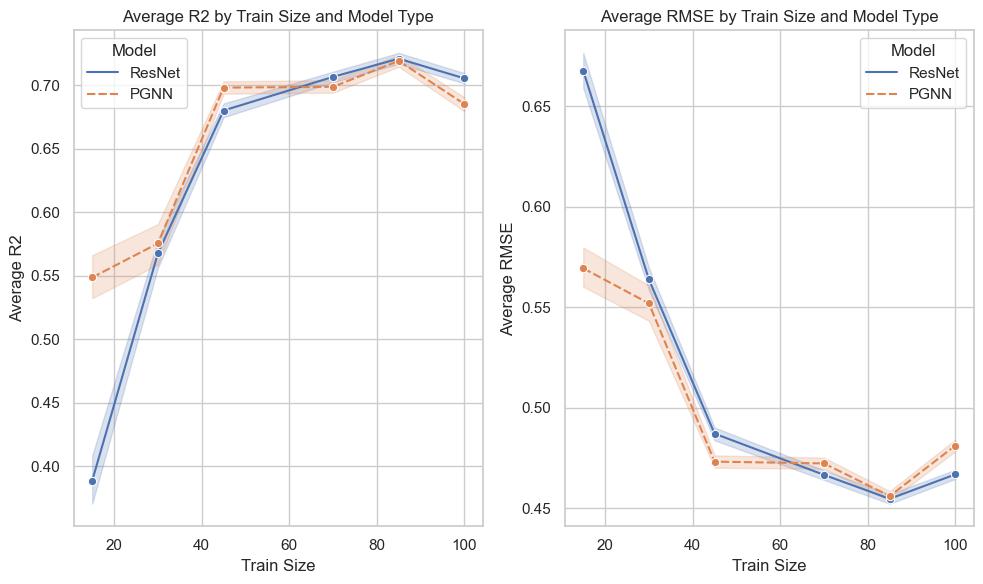}
\caption{Comparison of the performance metrics, Average $R^{2}$ and RMSE, between the PGNN and ResNet. The performance is evaluated across varying train sizes.}
    \label{Simulation1_a}
\end{figure}

\paragraph{Interpretation of Results}
The results indicate a consistent trend where the PGNN model generally outperforms the ResNet model, particularly at smaller training sizes. This is evident from the lower RMSE and MAE values and higher R\(^2\) scores for PGNN across most training sizes. The PGNN model's superiority in handling smaller datasets suggests its enhanced capability in learning from limited data, a crucial aspect in scenarios where collecting large datasets is challenging.

As the training size increases, both models show improvement in their performance metrics. However, the gap in performance between the two models narrows, indicating that ResNet benefits more significantly from larger training sets compared to PGNN.

\paragraph{Implications of the Study}
These results have significant implications for the application of neural network models in agricultural quality assessment. The robust performance of PGNN, especially with smaller datasets, underscores its potential in practical scenarios where data scarcity is a common challenge. Furthermore, the study highlights the importance of model selection based on the available data volume and the specific requirements of the quality assessment task.

\section{Application}
In this section, we delve deeper into the practical applications of our theoretical framework, focusing on two specific types of crops: cucumbers and mushrooms. Each  presents unique challenges and characteristics, making them ideal candidates to demonstrate the versatility and effectiveness of our proposed models. We will explore how our neural network architectures, specifically tailored for each case, perform in the task of quality assessment based on image analysis. 


Traditionally, image-based quality assessment in agriculture \cite{MEENU2021106} has often relied on colorimetric analysis, as seen in studies like the one by  \cite{mukherjee2022development}, which uses RGB and HSV color maps to classify mushrooms. These methods primarily focus on visible characteristics and are effective in scenarios where such attributes are strongly indicative of quality. Our PGNN-based approach extends beyond the surface-level features by integrating physical laws into the neural network's structure. This integration enables the model to capture not just the visual attributes, but also the underlying physical qualities that these attributes may represent. Such an approach is particularly advantageous in assessing quality attributes that are not immediately discernible from visual analysis alone.

By comparing our PGNN approach to traditional methods, we can see a shift from a purely data-driven, visual analysis to a more comprehensive, physics-guided assessment. This shift is indicative of the evolving landscape in agricultural technology, where the incorporation of domain-specific knowledge into machine learning models is becoming increasingly prevalent.

\subsection{Application to Cucumbers}
\paragraph{Dataset and Image Acquisition}
Our dataset consists of 660 RGB images of cucumbers, categorized as post-storage, with notable natural variations in color (from green to yellow) and texture. The images were preprocessed to standardize the Region of Interest (ROI) to \(110 \times 350\) pixels, followed by the application of Gaussian blur to refine the images. Image augmentation techniques, including resizing, random vertical and horizontal flips, and normalization, were employed to enhance the dataset's diversity.

\begin{figure}[ht]
\centering
\includegraphics[width=0.8\textwidth]{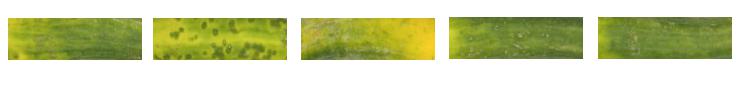}
\caption{Sample original cucumber images illustrating natural diversity.}
\label{fig:original_images}
\end{figure}

\begin{figure}[ht]
\centering
\includegraphics[width=0.8\textwidth]{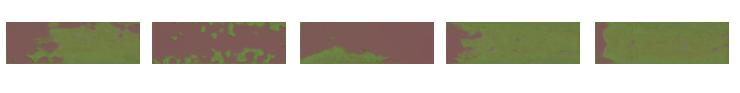}
\caption{Images processed to emphasize moisture content.}
\label{fig:inverse_regression_images}
\end{figure}

By utilizing inverse regression techniques, the process effectively enhances moisture-relevant attributes while suppressing irrelevant visual details. Consequently, the resulting images exhibit a more consistent representation, aligning with the PGNN's emphasis on analyzing moisture content. This transformation leads to a noticeable simplification in the images' color variance and textural detail when compared to their original counterparts, as illustrated in Figures \ref{fig:original_images} and \ref{fig:inverse_regression_images}. 

\paragraph{Label Description}
As described in \cite{Migal}, cucumber quality was labeled on a continuous scale from 0 to 17, based on laboratory measurements including elastic resistance, luminosity, change in weight, and size.

\paragraph{Model Training and Evaluation}
The models \( \hat{h} \) and \( \hat{f} \) were trained with a learning rate of 0.0001, while \( \hat{g}^{-1} \) was trained at a learning rate of 0.001. Training was conducted over 55 epochs, with results taken as the mean of epochs 25-55. Stratified splitting based on integer label values ensured balanced label distribution. The performance metrics used were RMSE, MAE, standard errors, and R\(^2\).

The dataset for this application is structured to allow a comprehensive evaluation of the model's performance and to understand the effect of training data size on its accuracy. Our dataset includes:

\begin{description}
  \item[Static Test Set:] This part of the dataset comprises a fixed test set consisting of 187 samples. It is used for evaluating the model's performance and serves as a standard benchmark for assessing the effectiveness of our machine learning model in predicting cucumber freshness.
  \item[Training Sets of Varying Sizes:] To explore the relationship between the size of the training set and model accuracy, training sets of different sizes are employed. These sets include 31,  47,  63,  79,  95, 111, 127, and 143 samples, allowing us to observe how the model's performance scales with an increase in training data.
\end{description}

\paragraph{Comparison Baselines and Objectives}
The study compares two models: Direct Prediction (\( \hat{h} \)) and Inverse Prediction (\( \hat{g}^{-1} \) and \( \hat{f} \)). The hypothesis posits that the Inverse Prediction model, particularly at smaller training sizes, outperforms the Direct Prediction method in accurately estimating cucumber quality.

\subsection{Results of the Cucumbers Data Application}

This section presents the findings from applying the Direct Prediction model (\( \hat{h} \), based on ResNet) and the combined Physics-Guided Inverse Regression $\hat{g}^{-1}$  and subsequent prediction model \( \hat{f} \) to real cucumber data. 

In the Fig. \ref{Cucumber8} below shown the performance of the \( \hat{h} \) (ResNet) and the combined \( \hat{g}^{-1} \) and \( \hat{f} \) (PGNN) models:

\begin{figure}[!ht]
\centering
\includegraphics[width=1\textwidth]{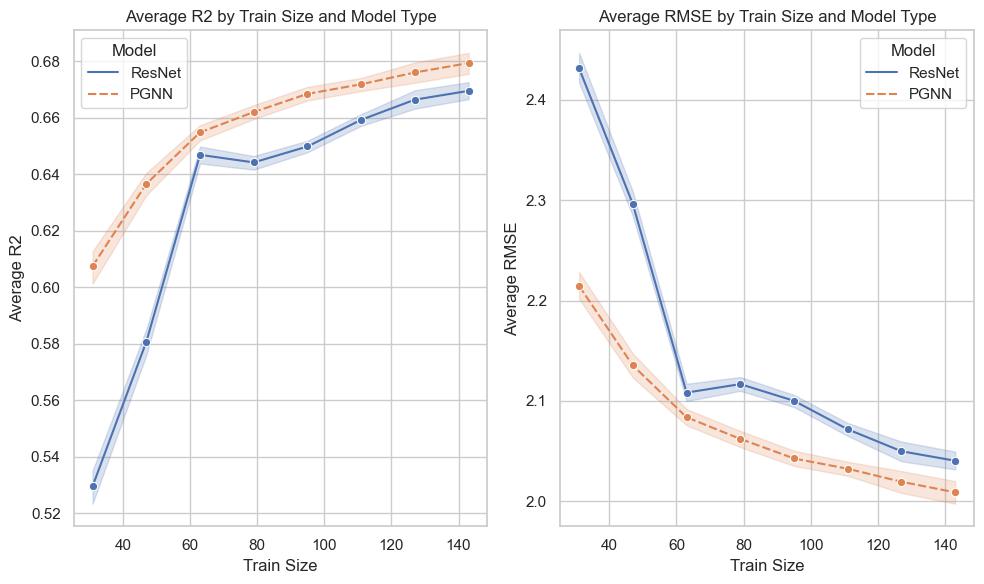}
\caption{Comparison of the performance metrics, Average $R^{2}$ and RMSE, between the PGNN and ResNet. The performance is evaluated across varying train sizes.}
    \label{Cucumber8}
\end{figure}

The findings indicate that the combined \( \hat{g}^{-1} \) and \( \hat{f} \) (PGNN) models generally show superior performance compared to the Direct Prediction model (\( \hat{h} \), ResNet), particularly in scenarios with smaller training sets. This trend is evident from the lower RMSE and MAE values and the higher R\(^2\) scores achieved by the PGNN models. These results support our hypothesis about the performance efficiency of PGNN models with varying training set sizes.

As training sizes increase, both models demonstrate improved performance, but the PGNN models maintain a consistent edge. This finding is particularly relevant in agricultural applications where data collection can be challenging, highlighting the effectiveness of PGNN models in scenarios with limited data availability.

The outcomes of this study underscore the potential of PGNN models in practical agricultural quality assessments, emphasizing their adaptability and robustness in various data environments. The superior performance of the PGNN models in handling real-world data variability and their efficiency in learning from smaller datasets make them a promising tool for agricultural imaging and analysis.

\subsection{Application to Mushrooms}
\paragraph{Dataset and Image Acquisition}
Our dataset comprises 332 images of mushrooms after post-harvest and cold storage, showcasing substantial variations in color and texture. These images underwent preprocessing to standardize the data for analysis, including ROI extraction and resizing to \(224 \times 224\) pixels.

\begin{figure}[ht]
\centering
\includegraphics[width=0.8\textwidth]{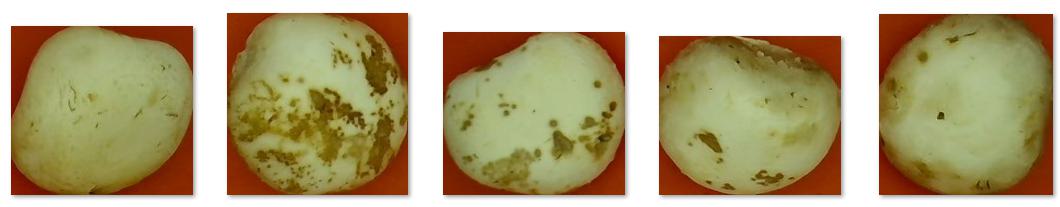}
\caption{Sample original mushrooms images illustrating natural diversity.}
\label{fig:mush_original_images}
\end{figure}

Figure \ref{fig:mush_original_images} displays the diversity in mushroom appearances, highlighting the challenge of automating quality assessment. This challenge is not only visual but also physical, as the quality of mushrooms is significantly influenced by their cellular structure and water holding capacity. Indeed, \cite{paudel2016effects} discuss how the water holding capacity, a key quality indicator, is affected by the cellular structure and the integrity of cell walls. Their findings emphasize the role of cell membrane integrity and the cell wall's structural components, such as chitin and proteins, in determining the mushroom's water holding capacity, especially under conditions like cold storage. This insight underscores the necessity of advanced image analysis techniques that can account for such intricate quality determinants, guiding our PGNN model to more accurately assess mushroom quality by considering these physical and biochemical factors.

\paragraph{Label Description}
The quality of mushrooms was quantified on a continuous scale from 0 to 5, integrating a multifaceted approach to capture the nuances of quality assessment based on chroma (color strength), luminosity (brightness), and additional specific properties as described in the associated labeling criteria. To enhance the precision of our quality assessment, we applied two specialized programs:

\begin{enumerate}
    \item The first program assesses the change in the L* value, which represents the lightness of the color, ranging from black (0) to white (100). This parameter is crucial for evaluating the color-based quality aspects of mushrooms, reflecting variations in freshness and overall appeal.
    \item The second program evaluates the fill ratio, analyzing the percentage of dark spots on the mushroom's surface. This measure is indicative of surface quality and is instrumental in detecting blemishes or imperfections that could affect the mushroom's visual and physical quality.
\end{enumerate}

These parameters, L* value and fill ratio, are integral to our comprehensive quality assessment model. By incorporating these measures into our evaluation, we ensure a more detailed and accurate representation of mushroom quality, capturing both color and surface condition aspects that are essential for determining the overall quality of the produce.


\paragraph{Preprocessing and Augmentation}
In addition to resizing and normalization, image augmentation techniques such as random vertical flip, random horizontal flip, and random rotation (up to 360 degrees) were applied to enhance the dataset's diversity and robustness for training.

\paragraph{Model Training and Evaluation}
Both the \( \hat{h} \) and \( \hat{f} \) models were trained with a learning rate of 0.0001, while the \( \hat{g}^{-1} \) model was trained at a higher rate of 0.001. Training was conducted over 55 epochs, with results averaged from epochs 25 to 55. Stratified splitting based on the integer value of labels ensured a balanced distribution across training sets. Performance metrics included RMSE, MAE, standard errors, and R\(^2\).

The dataset for this application is structured to allow a comprehensive evaluation of the model's performance and to understand the effect of training data size on its accuracy. Our dataset includes:

\begin{description}
  \item[Static Test Set:] This part of the dataset comprises a fixed test set consisting of 100 samples. It is used for evaluating the model's performance and serves as a standard benchmark for assessing the effectiveness of our machine learning model in predicting cucumber freshness.
  \item[Training Sets of Varying Sizes:] To explore the relationship between the size of the training set and model accuracy, training sets of different sizes are employed. These sets include 15,  31,  47,  63,  79,  95, 111, 127, and 143 samples, allowing us to observe how the model's performance scales with an increase in training data.
\end{description}
\paragraph{Comparison Baselines and Objectives}
The performance of two models is compared: Direct Prediction (\( \hat{h} \)) and Inverse Prediction (\( \hat{g}^{-1} \) and \( \hat{f} \)). The hypothesis is that, particularly for smaller training set sizes, the Inverse Prediction model will outperform the Direct Prediction method in accurately estimating mushroom quality.

\subsection{Results of the Mushroom Data Application}

The analysis of real mushroom data was conducted using the Direct Prediction model (\( \hat{h} \), based on ResNet) and the combined Physics-Guided Inverse Regression \( \hat{g}^{-1} \) with the subsequent prediction model \( \hat{f} \). 

In Fig. \ref{Shamp_Regr} shown the performance of the \( \hat{h} \) (ResNet) and the combined \( \hat{g}^{-1} \) and \( \hat{f} \) (PGNN) models on the mushroom dataset:

\begin{figure}[!ht]
\centering
\includegraphics[width=1\textwidth]{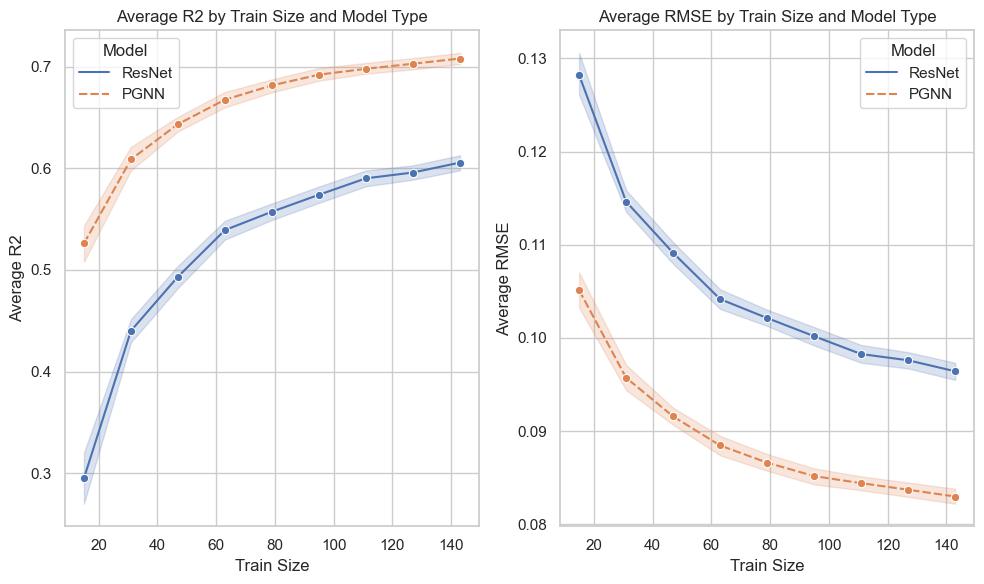}
\caption{Comparison of the performance metrics, Average $R^{2}$ and RMSE, between the PGNN and ResNet. The performance is evaluated across varying train sizes.}
    \label{Shamp_Regr}
\end{figure}

The results demonstrate a clear trend where the combined \( \hat{g}^{-1} \) and \( \hat{f} \) (PGNN) models outperform the Direct Prediction model (\( \hat{h} \), ResNet) across various training sizes, particularly in terms of RMSE and R\(^2\) scores. This trend is indicative of the PGNN model's enhanced capability in learning from limited data and its applicability in practical agricultural scenarios.

An improvement in performance metrics is observed for both models as the training size increases. However, the PGNN models maintain a consistent advantage, suggesting their robustness in learning from real-world data, particularly in the context of mushroom quality assessment.

As in the cucumber case, these findings underscore the potential of PGNN models in agricultural quality assessment, especially in scenarios where data collection is limited or challenging. The superior performance of the PGNN models in handling real-world data variability and their efficiency in learning from smaller datasets make them a promising tool for agricultural imaging and analysis.

\section{Summary and Discussion}

This study presented a comprehensive analysis of moisture content and crop
quality assessment in agricultural products using Physics-Guided Neural Networks (PGNNs). By integrating physical laws directly into the neural network architecture, we developed a model capable of accurately predicting moisture content from visual data. Our approach, grounded in the diffusion equation Eq. (\ref{diff_operator}), allowed for a nuanced understanding of moisture dynamics, demonstrating the PGNN model's ability to translate complex physical phenomena into actionable insights.

Our approach, characterized by the transformation network model \( \hat{g}^{-1} \), effectively mapped 3-channel image data \( Z \) to a representation of the estimated quality attributes \( \hat{x} \). The subsequent predictive modeling phase utilized these attributes to accurately predict laboratory measurements \( Y \), showcasing a significant improvement over traditional direct prediction models.

To validate the effectiveness of our two-stage PGNN method, we conducted comprehensive tests on two real-world datasets: Cucumbers and Mushrooms. These datasets, characterized by their distinct textural and morphological features, provided a robust foundation for assessing the predictive performance of our model across various training sizes. Our method consistently demonstrated superior performance compared to a direct prediction approach, particularly highlighting the advantage of the two-stage process in extracting and leveraging physics-guided features for predictive modeling.

To further substantiate our findings, we employed Monte Carlo simulations based on the simulated solution of the second Fick's law, a fundamental principle governing diffusion processes. These simulations, designed to mimic the transport phenomena underlying the datasets, provided a controlled environment to test the predictive capabilities of our method under known physical dynamics. While our PGNN method exhibited improved performance over the direct approach, the margin of enhancement was less pronounced compared to the real data tests.
The relatively smaller improvement observed in the Monte Carlo simulations can be explained by the nature of the simulated data, which, while governed by Fick's second law, lacked the complexity and noise inherent in real-world datasets. Consequently, the direct prediction approach, which relies solely on statistical correlations, performed comparably well in this less challenging scenario. This observation underscores the particular strength of our PGNN method in dealing with real, complex datasets where the integration of physical laws facilitates a deeper understanding and more accurate prediction of the target variables.

Future prospects include extending the PGNN framework to encompass a diverse array of produce types, thereby broadening its applicability and insights into various quality factors. Further exploration into a wider spectrum of physics-guided features, classification and multi-class classification and additional data types (e.g., hyperspectral) promises to deepen our understanding of crops quality assessment.

In conclusion, this research marks a significant milestone in the evolving landscape of physics-guided machine learning within the agricultural domain. The development of PGNN represents a step towards more advanced, precise, and reliable crops quality assessment systems. Such innovations are not merely academic; they have real-world implications, including the potential to elevate food quality standards, minimize waste, and improve economic outcomes across the agricultural industry.

\subsection*{Acknowledgement}

The author(s) disclosed receipt of the following financial support for the research, authorship, and/or publication of this article:

\begin{itemize}
  \item This project is supported by Israel Science Foundation grant (1755/22)
  \item This project is supported by Veterinary Services and Animal Health,
Ministry of Agriculture and Rural
Development
(21-06-0012)
\end{itemize}

\subsection*{Conflict of Interest}

The author has no conflicts to disclose.

\subsection*{Data Availability Statements}

The datasets generated and analyzed during this study can be obtained from the corresponding author upon reasonable request.

\bibliographystyle{unsrt}
\bibliography{references}  

\begin{thebibliography}{10}

\bibitem{kader2004role}
Adel~A Kader and Rosa~Sonya Rolle.
\newblock {\em The role of post-harvest management in assuring the quality and safety of horticultural produce}, volume 152.
\newblock Food \& Agriculture Org., 2004.

\bibitem{kiaya2014post}
Victor Kiaya.
\newblock Post-harvest losses and strategies to reduce them.
\newblock {\em Technical Paper on Postharvest Losses, Action Contre la Faim (ACF)}, 25:1--25, 2014.

\bibitem{rajapaksha2021reducing}
L~Rajapaksha, DMCC Gunathilake, SM~Pathirana, and T~Fernando.
\newblock Reducing post-harvest losses in fruits and vegetables for ensuring food security—case of sri lanka.
\newblock {\em MOJ Food Process Technols}, 9(1):7--16, 2021.

\bibitem{wognum2011systems}
PM~Nel Wognum, Harry Bremmers, Jacques~H Trienekens, Jack~GAJ Van Der~Vorst, and Jacqueline~M Bloemhof.
\newblock Systems for sustainability and transparency of food supply chains--current status and challenges.
\newblock {\em Advanced engineering informatics}, 25(1):65--76, 2011.

\bibitem{gomiero2018food}
Tiziano Gomiero.
\newblock Food quality assessment in organic vs. conventional agricultural produce: Findings and issues.
\newblock {\em Applied Soil Ecology}, 123:714--728, 2018.

\bibitem{fao2018future}
FAO.
\newblock The future of food and agriculture: alternative pathways to 2050.
\newblock {\em Food and Agriculture Organization of the United Nations Rome}, 2018.

\bibitem{spang2019food}
Edward~S Spang, Laura~C Moreno, Sara~A Pace, Yigal Achmon, Irwin Donis-Gonzalez, Wendi~A Gosliner, Madison~P Jablonski-Sheffield, Md~Abdul Momin, Tom~E Quested, Kiara~S Winans, et~al.
\newblock Food loss and waste: measurement, drivers, and solutions.
\newblock {\em Annual Review of Environment and Resources}, 44:117--156, 2019.

\bibitem{bhargava2021fruits}
Anuja Bhargava and Atul Bansal.
\newblock Fruits and vegetables quality evaluation using computer vision: A review.
\newblock {\em Journal of King Saud University-Computer and Information Sciences}, 33(3):243--257, 2021.

\bibitem{borras2015data}
Eva Borr{\`a}s, Joan Ferr{\'e}, Ricard Boqu{\'e}, Montserrat Mestres, Laura Ace{\~n}a, and Olga Busto.
\newblock Data fusion methodologies for food and beverage authentication and quality assessment--a review.
\newblock {\em Analytica Chimica Acta}, 891:1--14, 2015.

\bibitem{he2016deep}
Kaiming He, Xiangyu Zhang, Shaoqing Ren, and Jian Sun.
\newblock Deep residual learning for image recognition.
\newblock In {\em Proceedings of the IEEE conference on computer vision and pattern recognition}, pages 770--778, 2016.

\bibitem{lecun2015deep}
Yann LeCun, Yoshua Bengio, and Geoffrey Hinton.
\newblock Deep learning.
\newblock {\em Nature}, 521(7553):436--444, 2015.

\bibitem{goodfellow2016deep}
Ian Goodfellow, Yoshua Bengio, and Aaron Courville.
\newblock {\em Deep learning}.
\newblock MIT press, 2016.

\bibitem{krizhevsky2012imagenet}
Alex Krizhevsky, Ilya Sutskever, and Geoffrey~E Hinton.
\newblock Imagenet classification with deep convolutional neural networks.
\newblock {\em Advances in neural information processing systems}, 25, 2012.

\bibitem{cho2021determination}
Byeong-Hyo Cho and Shigenobu Koseki.
\newblock Determination of banana quality indices during the ripening process at different temperatures using smartphone images and an artificial neural network.
\newblock {\em Scientia Horticulturae}, 288:110382, 2021.

\bibitem{schmidt2020nonparametric}
Johannes Schmidt-Hieber.
\newblock {Nonparametric regression using deep neural networks with ReLU activation function}.
\newblock {\em The Annals of Statistics}, 48(4):1875 -- 1897, 2020.

\bibitem{kohler2022rate}
Michael Kohler, Adam Krzy{\.z}ak, and Benjamin Walter.
\newblock On the rate of convergence of image classifiers based on convolutional neural networks.
\newblock {\em Annals of the Institute of Statistical Mathematics}, 74(6):1085--1108, 2022.

\bibitem{langer2022statistical}
Sophie Langer and Johannes Schmidt-Hieber.
\newblock A statistical analysis of an image classification problem.
\newblock {\em arXiv preprint arXiv:2206.02151}, 2022.

\bibitem{kohler2021over}
Michael Kohler and Adam Krzy{\.z}ak.
\newblock Over-parametrized deep neural networks minimizing the empirical risk do not generalize well.
\newblock {\em Bernoulli}, 27(4):2564--2597, 2021.

\bibitem{raissi2019physics}
Maziar Raissi, Paris Perdikaris, and George~E Karniadakis.
\newblock Physics-informed neural networks: A deep learning framework for solving forward and inverse problems involving nonlinear partial differential equations.
\newblock {\em Journal of Computational physics}, 378:686--707, 2019.

\bibitem{karniadakis2021physics}
George~Em Karniadakis, Ioannis~G Kevrekidis, Lu~Lu, Paris Perdikaris, Sifan Wang, and Liu Yang.
\newblock Physics-informed machine learning.
\newblock {\em Nature Reviews Physics}, 3(6):422--440, 2021.

\bibitem{cuomo2022scientific}
Salvatore Cuomo, Vincenzo~Schiano Di~Cola, Fabio Giampaolo, Gianluigi Rozza, Maziar Raissi, and Francesco Piccialli.
\newblock Scientific machine learning through physics--informed neural networks: Where we are and what’s next.
\newblock {\em Journal of Scientific Computing}, 92(3):88, 2022.

\bibitem{karpatne2017theory}
Anuj Karpatne, Gowtham Atluri, James~H Faghmous, Michael Steinbach, Arindam Banerjee, Auroop Ganguly, Shashi Shekhar, Nagiza Samatova, and Vipin Kumar.
\newblock Theory-guided data science: A new paradigm for scientific discovery from data.
\newblock {\em IEEE Transactions on knowledge and data engineering}, 29(10):2318--2331, 2017.

\bibitem{daw2022physics}
Arka Daw, Anuj Karpatne, William~D Watkins, Jordan~S Read, and Vipin Kumar.
\newblock Physics-guided neural networks (pgnn): An application in lake temperature modeling.
\newblock In {\em Knowledge Guided Machine Learning}, pages 353--372. Chapman and Hall/CRC, 2022.

\bibitem{9126036}
Lei Wang, Qun Zhou, and Shuangshuang Jin.
\newblock Physics-guided deep learning for power system state estimation.
\newblock {\em Journal of Modern Power Systems and Clean Energy}, 8(4):607--615, 2020.

\bibitem{ongie2020deep}
Gregory Ongie, Ajil Jalal, Christopher~A Metzler, Richard~G Baraniuk, Alexandros~G Dimakis, and Rebecca Willett.
\newblock Deep learning techniques for inverse problems in imaging.
\newblock {\em IEEE Journal on Selected Areas in Information Theory}, 1(1):39--56, 2020.

\bibitem{yaman2020self}
Burhaneddin Yaman, Seyed Amir~Hossein Hosseini, Steen Moeller, Jutta Ellermann, K{\^a}mil U{\u{g}}urbil, and Mehmet Ak{\c{c}}akaya.
\newblock Self-supervised learning of physics-guided reconstruction neural networks without fully sampled reference data.
\newblock {\em Magnetic resonance in medicine}, 84(6):3172--3191, 2020.

\bibitem{fieguth2010statistical}
Paul Fieguth.
\newblock {\em Statistical image processing and multidimensional modeling}.
\newblock Springer Science \& Business Media, 2010.

\bibitem{rawat2017deep}
Waseem Rawat and Zenghui Wang.
\newblock Deep convolutional neural networks for image classification: A comprehensive review.
\newblock {\em Neural computation}, 29(9):2352--2449, 2017.

\bibitem{cortbaoui2015new}
Patrick~E. Cortbaoui and Michael~O. Ngadi.
\newblock New method to quantify postharvest quality loss of cucumber using the taguchi approach.
\newblock {\em Food Science and Quality Management}, 44:13--22, 2015.

\bibitem{poenicke1977ethylene}
EF~Poenicke, SJ~Kays, DA~Smittle, and RE~Williamson.
\newblock Ethylene in relation to postharvest quality deterioration in processing cucumbers1.
\newblock {\em Journal of the American Society for Horticultural Science}, 102(3):303--306, 1977.

\bibitem{manjunatha2014effect}
M~Manjunatha and Rahul~K Anurag.
\newblock Effect of modified atmosphere packaging and storage conditions on quality characteristics of cucumber.
\newblock {\em Journal of food science and technology}, 51:3470--3475, 2014.

\bibitem{nasef2018short}
Ibrahim~N Nasef.
\newblock Short hot water as safe treatment induces chilling tolerance and antioxidant enzymes, prevents decay and maintains quality of cold-stored cucumbers.
\newblock {\em Postharvest Biology and Technology}, 138:1--10, 2018.

\bibitem{cantor2020equations}
Brian Cantor.
\newblock {\em The equations of materials}.
\newblock Oxford University Press, 2020.

\bibitem{kahramanouglu2019improving}
{\.I}brahim Kahramano{\u{g}}lu and Serhat Usanmaz.
\newblock Improving postharvest storage quality of cucumber fruit by modified atmosphere packaging and biomaterials.
\newblock {\em HortScience}, 54(11):2005--2014, 2019.

\bibitem{omolola2017quality}
Adewale~O Omolola, Afam~IO Jideani, and Patrick~F Kapila.
\newblock Quality properties of fruits as affected by drying operation.
\newblock {\em Critical reviews in food science and nutrition}, 57(1):95--108, 2017.

\bibitem{yahia2018postharvest}
Elhadi~M Yahia and Armando Carrillo-Lopez.
\newblock {\em Postharvest physiology and biochemistry of fruits and vegetables}.
\newblock Woodhead publishing, 2018.

\bibitem{onwude2016modeling}
Daniel~I Onwude, Norhashila Hashim, Rimfiel~B Janius, Nazmi~Mat Nawi, and Khalina Abdan.
\newblock Modeling the thin-layer drying of fruits and vegetables: A review.
\newblock {\em Comprehensive reviews in food science and food safety}, 15(3):599--618, 2016.

\bibitem{metropolis1949monte}
Nicholas Metropolis and Stanislaw Ulam.
\newblock The monte carlo method.
\newblock {\em Journal of the American statistical association}, 44(247):335--341, 1949.

\bibitem{bracewell1966fourier}
Ron Bracewell and Peter~B Kahn.
\newblock The fourier transform and its applications.
\newblock {\em American Journal of Physics}, 34(8):712--712, 1966.

\bibitem{MEENU2021106}
Maninder Meenu, Chinmay Kurade, Bala~Chakravarthy Neelapu, Sahil Kalra, Hosahalli~S. Ramaswamy, and Yong Yu.
\newblock A concise review on food quality assessment using digital image processing.
\newblock {\em Trends in Food Science \& Technology}, 118:106--124, 2021.

\bibitem{mukherjee2022development}
Alok Mukherjee, Tanmay Sarkar, Kingshuk Chatterjee, Dibyajit Lahiri, Moupriya Nag, Maksim Rebezov, Mohammad~Ali Shariati, Alevtin Miftakhutdinov, and Jose~M Lorenzo.
\newblock Development of artificial vision system for quality assessment of oyster mushrooms.
\newblock {\em Food Analytical Methods}, 15(6):1663--1676, 2022.

\bibitem{Migal}
Ofer~M. Shir, Boris Yazmir, Assaf Israeli, and Dan Gamrasni.
\newblock Algorithmically-guided postharvest protocols by experimental combinatorial optimization.
\newblock In {\em Proceedings of the Genetic and Evolutionary Computation Conference Companion}, GECCO '22, page 2027–2035, New York, NY, USA, 2022. Association for Computing Machinery.

\bibitem{paudel2016effects}
Ekaraj Paudel, Remko~M Boom, Els van Haaren, Joanne Siccama, and Ruud~GM van~der Sman.
\newblock Effects of cellular structure and cell wall components on water holding capacity of mushrooms.
\newblock {\em Journal of Food Engineering}, 187:106--113, 2016.

\end{thebibliography}

\pagebreak

\end{document}